\documentclass[preprint2]{aastex}

\usepackage{color}

\slugcomment{Resubmitted to Astrophysical Journal in February 2016}

\shorttitle{Stable and Unstable Regimes of Mass Accretion onto RW Aur A}
\shortauthors{Takami et al.}

\begin{document}



\title{Stable and Unstable Regimes of Mass Accretion onto RW Aur A}


\author{
Michihiro Takami\altaffilmark{1},
Yu-Jie Wei\altaffilmark{1},
Mei-Yin Chou\altaffilmark{1},
Jennifer L. Karr\altaffilmark{1},
Tracy L. Beck\altaffilmark{2},
Nadine Manset\altaffilmark{3},
Wen-Ping Chen\altaffilmark{4},
Ryuichi Kurosawa\altaffilmark{5},
Misato Fukagawa\altaffilmark{6},
Marc White\altaffilmark{7},
Roberto Galv\'an-Madrid\altaffilmark{8},
Hauyu Baobab Liu\altaffilmark{1,9},
Tae-Soo Pyo\altaffilmark{10},
Jean-Francois Donati\altaffilmark{11}
}
\altaffiltext{1}{Institute of Astronomy and Astrophysics, Academia Sinica, 11F of Astronomy-Mathematics Building, AS/NTU, No.1, Sec. 4, Roosevelt Rd, Taipei 10617, Taiwan, R.O.C.} 
\altaffiltext{2}{The Space Telescope Science Institute, 3700 San Martin Dr. Baltimore, MD 21218, USA}
\altaffiltext{3}{Canada-France-Hawaii Telescope, 65-1238 Mamalahoa Hwy, Kamuela, HI 96743, USA}
\altaffiltext{4}{Institute of Astronomy, National Central University, Taiwan 320, Taiwan}
\altaffiltext{5}{Max-Planck-Institut f\"ur Radioastronomie, Auf dem H\"ugel 69, D-53121 Bonn, Germany}
\altaffiltext{6}{National Astronomical Observatory of Japan, 2-21-1 Osawa, Mitaka, Tokyo 181-8588, Japan}
\altaffiltext{7}{Research School of Astronomy \& Astrophysics, The Australian National University, Cotter Rd., Weston, ACT, Australia, 2611}
\altaffiltext{8}{Centro de Radioastronom\'ia y Astrof\'isica, UNAM, Apdo. Postal 3-72 (Xangari), 58089 Morelia, Michoac\'an, M\'exico}
\altaffiltext{9}{European Southern Observatory, Karl-Schwarzschild-Str. 2, D-85748 Garching, Germany}
\altaffiltext{10}{Subaru Telescope, 650 North Aohoku Place, Hilo, HI 96720, USA}
\altaffiltext{11}{UPS-Toulouse/CNRS-INSU, Institut de Recherche en Astrophysique et Pla\'netologie (IRAP) UMR 5277, Toulouse F-31400, France}


\begin{abstract}
We present monitoring observations of the active T Tauri star RW Aur, from 2010 October to 2015 January, using optical high-resolution ($R \ge 10000$) spectroscopy with CFHT-ESPaDOnS. Optical photometry in the literature shows bright, stable fluxes over most of this period, with lower fluxes (by 2-3 mag.) in 2010 and 2014. In the bright period our spectra show clear photospheric absorption, complicated variation in the \ion{Ca}{+2} 8542 \AA~{ emission} profile shapes, and a large variation in redshifted absorption in the \ion{O}{+1} 7772 and 8446 \AA~and \ion{He}{+1} 5876 \AA~lines, suggesting unstable mass accretion during this period. In contrast, these line profiles are relatively uniform during the faint periods, suggesting stable mass accretion. { During the faint periods the photospheric absorption lines are absent or marginal, and the averaged \ion{Li}{1} profile shows redshifted absorption due to an inflow.
We discuss (1) occultation by circumstellar material or a companion and (2) changes in the activity of mass accretion to explain the above results, together with near-infrared and X-ray observations from 2011-2015. Neither scenario can simply explain all the observed trends, and more theoretical work is needed to further investigate their feasibilities.}
\end{abstract}




\section{INTRODUCTION}

Understanding the mechanisms for mass accretion and ejection is one of the key issues in star formation theories. Low-mass pre-main sequence stars, the ``T-Tauri stars'' have been extensively studied to address this issue. The optically visible nature of these objects has allowed astronomers to extensively observe them at optical and and near-infrared wavelengths over the years.

The current paradigm of mass accretion based on these observations has been summarized by, e.g., \citet[][]{Calvet00,Najita00}. The stellar magnetospheres of accreting T Tauri stars (classical T Tauri stars, hereafter CTTSs) are associated with the inner edges of the circumsteller disks, and they regulate the stellar rotation. Mass accretion from the disk to the star occurs through this magnetic field. These accretion flows, called magnetospheric accretion flows, are associated with optical and near-infared permitted line emission, in particular with the broad component (a full width half maximum velocity $V_{\rm FWHM} > 100$ km s$^{-1}$). Accretion shocks at the stellar photosphere cause hot spots, with a typical temperature of $\sim 10^4$ K \citep[see][for the measurement of temperatures]{Gullbring98}, and these add blue excess continuum to the photospheric emission ($T_{\rm eff} \sim 4000$ K). The blue excess continuum does not show photospheric absorption, and as a result, it makes a number of photospheric absorption lines shallower (``veiling''). The luminosities of the blue excess continuum and permitted lines monotonically increase with the mass accretion rate.

Optical continuum and emission lines show variabilities on different time scales. A periodic variability of 2--20 days is in most cases well explained by a combination of stellar rotation and the non-axisymmetric distribution of hot spots and magnetospheric accretion columns \citep[][for reviews]{Bouvier07_PPV,Bouvier14}, while time-variable mass accretion in close binaries explains such a variability in DQ Tau \citep{Mathieu97}. Furthermore, the following mechanisms seem to cause irregular variabilities over wide ranges of time scales: (1) time-variable mass accretion; (2) magnetic flaring activities; and (3) obscuration by circumstellar dust \citep[e.g.,][]{Herbst94}. Extensive studies of variabilities could lead us to better understand the physics and geometry of stellar magnetospheres and magnetospheric mass accretion. The stability of magnetic structures at the star and the inner disk region may also affect the architecture of planetary systems such as the orbits of hot Jupiters \citep[e.g.,][]{Lin96,Romanova06,Papaloizou07}.

As for other young stellar objects, many CTTSs are also associated with a collimated jet, which is often observed in optical and near-IR forbidden line emission. It is believed that these jets are powered by mass accretion. Theories predict that the jets are launched either at the inner edge of the disk, or from a wider disk surface. See, e.g., \citet{Eisloffel00,Ray07,Frank14} for reviews. In the former case, the activities of the stellar magnetosphere and magnetospheric mass accretion described above may also affect the mass ejection seen in the jet.

RW Aur A is one of the best studied CTTSs.
Its stellar properties are summarized in Table \ref{table:RWAur}.
The star shows a variety of emission lines, indicating active accretion \citep[e.g.,][]{Hamann92a,Muzerolle98_opt}. These lines show a complicated variability, which has been extensively studied by \citet{Gahm99,Petrov01a,Alencar05}. 
The star also hosts a known companion with a separation of 1\farcs4--1\farcs5
\citep[RW Aur B; e.g., ][]{Bisikalo12}.
In addition to this known companion, RW Aur A itself is a suspected spectroscopic binary \citep{Hartmann86,Gahm99,Petrov01b}.
RW Aur A is also associated with a pair of asymmetric extended jets, which have also been extensively studied \citep[e.g.,][]{Lopez03,Pyo06,Beck08,Melnikov09,Liu12,Coffey12}.

Optical photometry of the RW Aur system shows irregular variabilities, ranging from days to years \citep[][]{Gahm93,Herbst94,Gahm99,Beck01,Petrov01a,Petrov01b,Alencar05,Grankin07}.
%
\citet{Rodriguez13,Antipin15,Petrov15} reported remarkable decreases in optical photometric fluxes in 2010 and 2014, by 2--3 mag. 
The optical colors did not change during this period, consistent with the explanation that the flux decreases are due to gray extinction. These authors point out the possibility that the gray extinction is due to occultation of the star by a tidally disrupted dusty disk or a wind with large dust grains.

We report an intriguing variability of optical high-resolution spectra during the above period. Different behaviors in the variations between the bright and faint periods indicate different activities of mass accretion. In Section 2, we describe the observations. In Section 3 we show our results for the continuum and line emission. In Section 4 { we attempt to explain the observations together with X-ray to near-infrared observations in the literature.}
In Section 5 we summarize our results and conclusions.


\section{OBSERVATIONS AND DATA REDUCTION}

The observations were made using the 3.6-m Canada-France-Hawaii Telescope (CFHT) with ESPaDOnS, covering the wavelength range 3700--10500 \AA. The spectra of RW Aur A were obtained using  the ``object+sky spectroscopy mode", for which the spectra of the object and sky are simultaneously obtained through different fibers. This mode provides a spectral resolution of 68,000.

Table \ref{table:log} shows the log of the observations. All the dates described in the paper are based on UT. The observations were made from 2010 October to 2015 January in five autumn-winter semesters (August to January, hereafter 2010B--2014B), with 2--4 observing runs for each semester, and 1--7 visits during each run, with intervals of 1--10 nights.
 This is a part of our long-term monitoring of a few active pre-main sequence stars (RW Aur, DG Tau, RY Tau, XZ Tau), for which the emission line spectra obtained in 2010 were published in \citet{Chou13}. Seeing was 0\farcs8 or less for about 70 \% of the visits, and it exceeded 1\arcsec~ during a few visits, reaching up to 1\farcs3. 
During each visit two telluric standards (HD 42784/283642) were observed at a similar airmass.

All spectra were made with the queue mode at the observatory.  A single 360-s exposure was made for most of the nights. The constancy of the exposures over the entire period causes systematic differences in signal-to-noise between 2010B, 2014B and the other semesters as shown in later sections, because of the large flux changes described in Section 1. Two 360-s exposures with similar signal-to-noise were obtained for a few nights (see Table \ref{table:log}) with an interval of 6-13 min. These two spectra were nearly identical to each other, and we averaged the spectra for each night for better signal-to-noise. 

{ While spectrophotometry was successful in 2010B \citep{Chou13}, it was not successful for some observing runs, probably due to the presence of cirrus or patchy clouds during the observations. Therefore, our measurements and discussion will be focused on data without absolute fluxes, including emission and absorption features normalized to the continuum, equivalent widths, and the color of the continuum.

The instrument aperture (1\farcs6 in diameter) was centered on RW Aur A, to avoid flux from RW Aur B (1\farcs4 from the star).
Even so, a few spectra obtained in the faint periods (i.e., 2010B and 2014B) are suspected to be contaminated by the flux from RW Aur B. We excluded these spectra from our analysis. Photospheric absorption spectra obtained on a few nights are also suspected to be contaminated by this secondary star, which appears to have significantly larger absorption than the primary star\footnote{ We obtained a spectrum of B on 2016 January 16 and found significantly larger photospheric absorption for RW Aur B than any RW Aur A spectra in our study.} (Section 3.1).
}

In addition to RW Aur A, we use ESPaDOnS spectra of four weak-lined T Tauri stars (WTTSs), TaP 4 (K1), Par 1379 (K4), V2129 Oph (K5) and RXJ 1608.0-3857 (M0). The observations were made in 2013--2014 as a part of a large survey of WTTSs, ``The Magnetic Topologies of Young Stars and the Survival of close-in giant Exoplanets" (MaTYSSE) programme \citep[e.g.,][]{Donati14,Donati15}, with the spectropolarimetry mode and a spectral resolution of 65,000. Their Stokes $I$ spectra will be used to compare stellar absorption with RW Aur A and for veiling measurements.

Data were reduced using the standard pipeline ``Upena'' provided by the CFHT, which is based on the Libre-ESpRIT package \citep{Donati97}. Sky subtraction has been performed for the RW Aur spectra. Although sky spectra were not obtained for the WTTSs, the contribution of sky emission to these spectra is negligible for the wavelength coverages used below. Telluric absorption was corrected for the [\ion{O}{+1}] 6300 \AA~line. The contribution of telluric absorption is negligible for the other emission lines and continuum spectra discussed below.


\section{RESULTS}

In Section 3.1 we show continuum spectra at 5900--6130 \AA, and analysis together with that at 5720--5860 \AA~and 8630--8640 \AA.
These wavelength ranges are selected for the following reasons: (1) high signal-to-noise; (2) absence of emission lines, which cover a significant fraction of the entire spectra; and (3) absence of telluric absorption lines for simpler data reduction. We also show the continuum flux ratios between 5750--5850 and 8630--8640 \AA~(hereafter $F_{\lambda 5800}/F_{\lambda 8635}$) {, and the \ion{Li}{1} 6708 \AA~absorption profiles}.
%
%

In Section 3.2 we show line profiles and equivalent widths of H$\alpha$ 6563 \AA, \ion{Ca}{+2} 8542 \AA, \ion{O}{+1} 7772 \AA~and 8446 \AA, \ion{He}{+1} 5876 \AA, and [\ion{O}{+1}] 6300 \AA~ lines. This paper will be focused on the different behaviors of time variations between the photometrically bright and faint periods. A more detailed study of emission lines will be provided in a forthcoming paper.


\subsection{Continuum}
\begin{figure*}
\epsscale{2.3}
\plotone{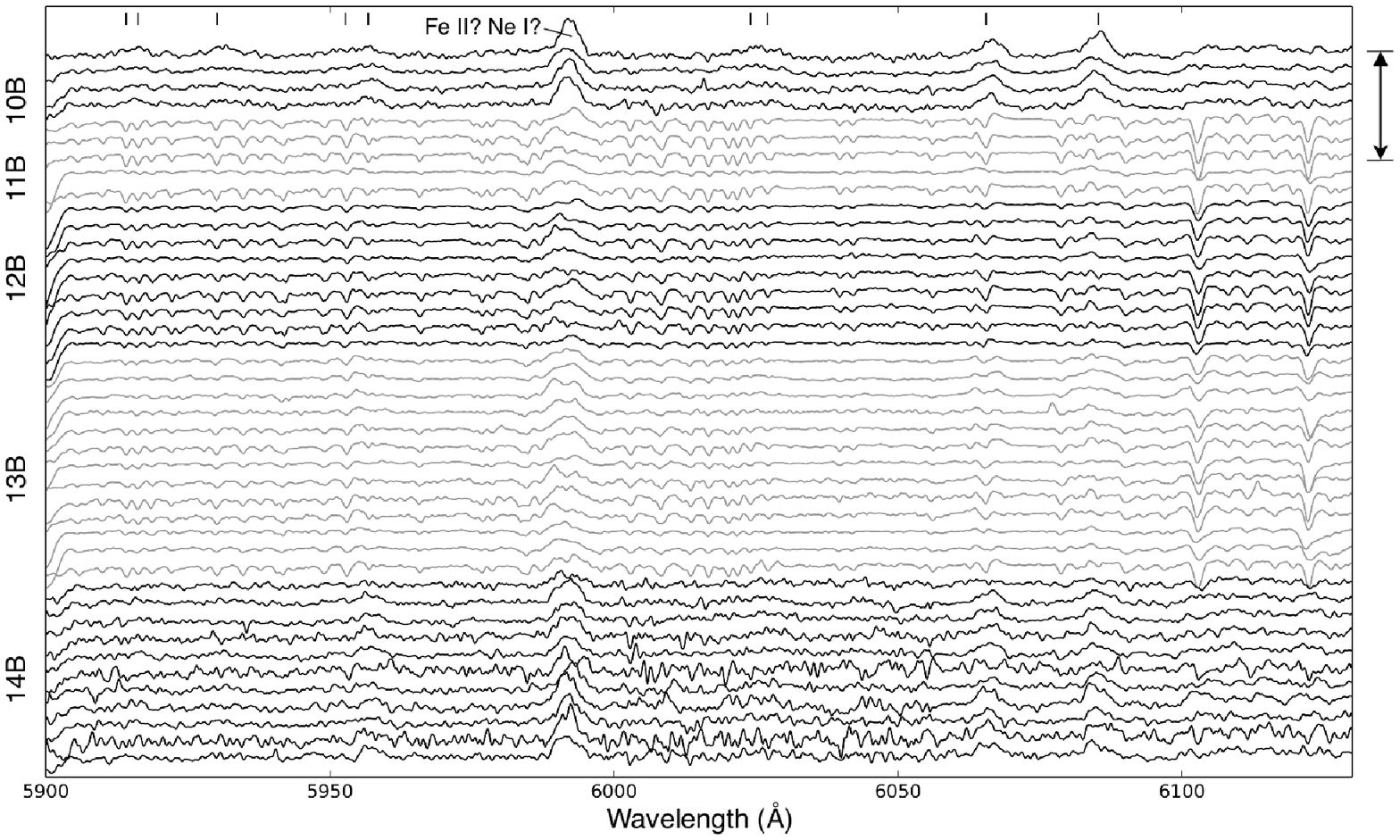}
\caption{The continuum spectra of RW Aur A at 5900--6130 \AA.  
The black and gray colors are used to identify spectra observed in different semesters. 
Each spectrum is normalized to the continuum and convolved with a Gaussian with a FWHM of 30 km s$^{-1}$. The continuum level of each spectrum is shown with an arrow at the top right.
The spectra observed on 2010 October 16 { and} 21 are not plotted due to modest signal-to-noise.
The ticks at the top are the wavelengths of \ion{Fe}{+1} lines probably corresponding to marginal emission features in 2010B and 2014B.
\label{fig:continuum}}
\end{figure*}

\begin{figure}
\epsscale{1.1}
\plotone{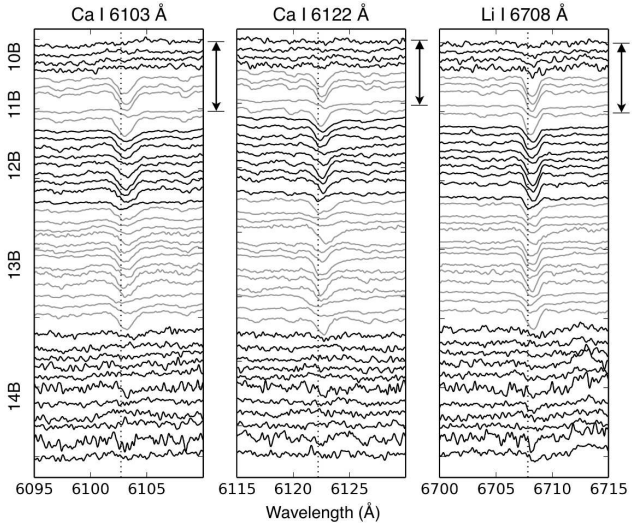}
\caption{ Same as Figure \ref{fig:continuum} but for the \ion{Ca}{1} 6103/6122 \AA~and  \ion{Li}{1} 6708 \AA~lines. Each profile is convolved with a Gaussian with a FWHM of 10 km s$^{-1}$. The dotted lines show the rest velocities of individual transitions.
\label{fig:CaI_LiI}}
\end{figure}

Figure \ref{fig:continuum} shows the spectra at 5900--6130 \AA.
{ Figure \ref{fig:CaI_LiI} shows the \ion{Ca}{1} 6103/6122 \AA~lines, the clearest absorption lines in Figure \ref{fig:continuum}. In Figure  \ref{fig:CaI_LiI} we also show the \ion{Li}{1} 6708 \AA~absorption line, one of the clearest absorption lines in the entire spectrum. The spectra in Figures \ref{fig:continuum} and \ref{fig:CaI_LiI} are convolved using Gaussians with FWHM velocities of 30 and 10 km s$^{-1}$, respectively, to increase signal-to-noise. A larger FWHM is used for the spectra in Figure \ref{fig:continuum} to best show shallow absorption lines.}


{ In Figure \ref{fig:continuum} the spectra in 2011B--2013B show a number of photospheric absorption lines}.
We attempt to measure the veiling for these spectra using the following equation:
\begin{equation}
r (\lambda) = F_{\rm veil} (\lambda) / F_{\rm phot} (\lambda)
\end{equation}
where $F_{\rm veil} (\lambda)$ and $F_{\rm phot} (\lambda)$ are fluxes from the blue excess continuum and stellar photosphere, respectively \citep[e.g.,][]{Muzerolle98_opt}. As described in Section 1, we assume that $F_{\rm veil} (\lambda)$ does not have any absorption features. We use a spectrum of a WTTS for $F_{\rm phot} (\lambda)$ as in \citet{Hartigan95,Hartigan03,Muzerolle98_opt}. The veiling $r$ is then determined using Equation (1) and the least squares method.

We found that, of the four WTTSs, Par 1379 (K4) and V2129 Oph (K5) provide the least variance. Differences in variance are marginal between the use of these two stars, hence we derive the veiling of each RW Aur spectrum using these two stars separately, then average the values. We note that the veiling obtained using Par 1379 is systematically larger than V2129 Oph by typically 5--10 \%.  The spectral type of K4--K5 is consistent with previous measurements of the spectral type of K1--K4 \citep{Hartigan95,Muzerolle98_opt,White01,Petrov01a}.

Table \ref{table:veiling} shows the veiling $r$ measured at 5720--5860 \AA~and 6000--6130 \AA. The values over the two wavelength ranges agree well. Hereafter we average $r$ measured over these two wavelength ranges and refer to it as the veiling at 6000 \AA~or $r_{\lambda 6000}$. The veiling $r_{\lambda 6000}$ ranges from 0.7--5, similar to that measured from 1996--1999 at 5560--6050 \AA~by \citet{Petrov01a}. The measured range corresponds to a change in the total flux of up to a factor of 3 with Equation (1) and the assumptions described above. This change in total flux agrees with the photometric variation reported by \citet{Rodriguez13,Petrov15} ($\sim$1 mag., corresponding a factor of $\sim$2.5 in flux). 

{ In Figures \ref{fig:continuum} and \ref{fig:CaI_LiI} photospheric absorption is absent in most of the spectra observed in 2010B and 2014B. This is in contrast with the remarkable absorption observed in 2011B--2013B. The absence of photospheric absorption in 2010B and 2014B may be partially due to the fact that the spectra are covered by emission lines. In Figures \ref{fig:continuum} we find several marginal and broad \ion{Fe}{1} emission lines in addition to the \ion{Fe}{2} or \ion{Ne}{1} emission at $\sim$5990 \AA.

A few spectra observed in 2010B and 2014B show marginal absorption in Figure  \ref{fig:CaI_LiI}. It is not clear if these absorption lines are associated with the target star RW Aur A, or if they are due to contaminating emission from RW Aur B (Section 2). The primary star was significantly fainter during these periods \citep{Antipin15}, which would have allowed contaminating emission from RW Aur B to be seen more clearly in the observed spectra.}

\begin{figure}
\epsscale{0.8}
\plotone{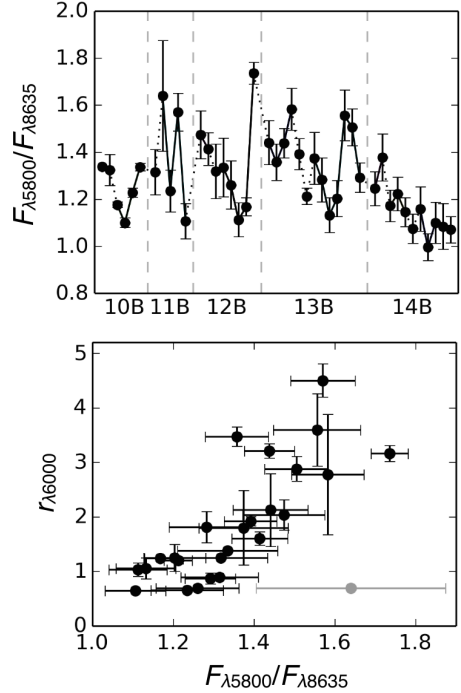}
\caption{($upper$) Time variation of the $F_{\lambda 5800}/F_{\lambda 8635}$ continuum flux ratio. The values for different visits are plotted with a constant interval along the horizontal axis regardless of the actual time intervals. The values observed in the same observing run are connected using sold lines. Those for different observing run are connected using dotted lines. The uncertainty of the measurement is primarily due to slightly different values obtained using two standard stars (HD 283642 and HD 42784) for calibration. ($lower$) Correlation between the $F_{\lambda 5800}/F_{\lambda 8635}$ flux ratio and veiling in 2011B--2013B. The gray dot is the data point for 2012 January 5, which is offset from a correlation between the veiling and the flux ratio perhaps due to a large uncertainty in $F_{\lambda 5800}/F_{\lambda 8635}$. The uncertainty of the veiling is determined by different values at 5720--5860 \AA~and 6000--6130 \AA~(see text). 
\label{fig:cont_f}}
\end{figure}

Figure \ref{fig:cont_f} shows the time variation of the $F_{\lambda 5800}/F_{\lambda 8635}$ continuum flux ratios and their correlation with veiling. { When selecting these wavelengths we considered maximizing the color variation with respect to systematic errors, and avoided deep absorption lines in the spectra of the standard stars.}
The flux ratios are similar in 2010B--2013B, while those in 2014B are systematically lower.
A similar trend was also observed by \citet{Petrov15}, who reported that the $V-R$ magnitude was larger at the photometric minimum in 2014, but did not show a clear difference between the minimum in 2010 and the bright state in 2011--2013. In Figure \ref{fig:cont_f} the $F_{\lambda 5800}/F_{\lambda 8635}$ flux ratio in 2011B--2013B shows a good positive correlation with the veiling, except for the data from 2012 January 5, which has a large uncertainty. This implies that the excess continuum is ``hotter'' than the stellar photosphere, agreeing with present veiling theory \citep[e.g.,][for a review]{Calvet00}.

\begin{figure*}
\epsscale{2.1}
\plotone{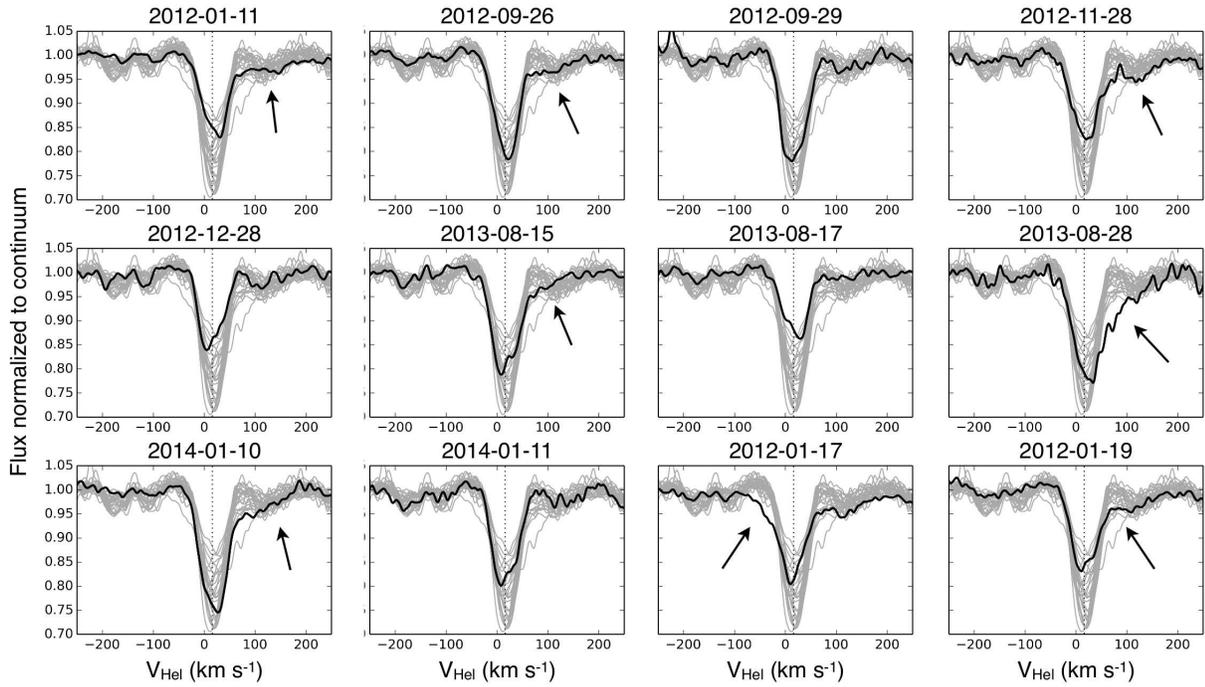}
\caption{ Asymmetric \ion{Li}{1} 6708 \AA~profiles observed from 2011B--2013B (black). Gray profiles in the individual boxes are all the profiles observed from 2011B--2013B. These profiles are convolved with a Gaussian with a FWHM of 10 km s$^{-1}$. The arrows show the blueshifted and redshifted wing absorption discussed in the text. The dotted lines show the median velocity of the line ($V_{\rm Hel}$=17 km s$^{-1}$, see text).
\label{fig:Li_asymmetries}}
\end{figure*}

\begin{figure}
\epsscale{0.9}
\plotone{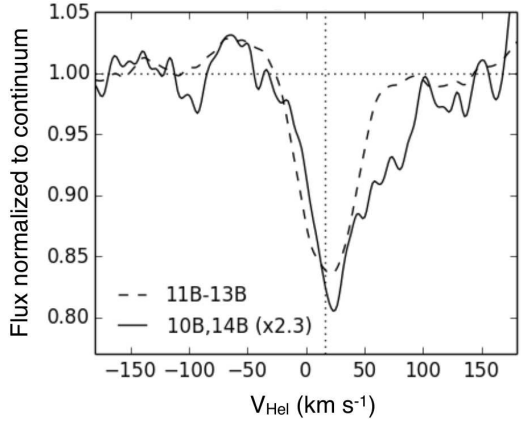}
\caption{ The \ion{Li}{1} 6708 \AA~profiles averaged over 2011B-2013B (dashed curve) and 2010B+2014B (solid curve). Each spectrum was normalized to the continuum, and  convolved with a Gaussian with a FWHM of 10 km s$^{-1}$ before averaging. The absorption and possible emission features in the 2010B+2014B profile are magnified by a factor of 2.3 for the best comparison with the 2011B-2013B profile.  The dotted vertical line shows a median velocity of the line ($V_{\rm Hel}$=17 km s$^{-1}$, see text), and the dotted horizontal line shows the continuum level. 
\label{fig:Li_averaged_profiles}}
\end{figure}

\begin{figure}
\epsscale{0.9}
\plotone{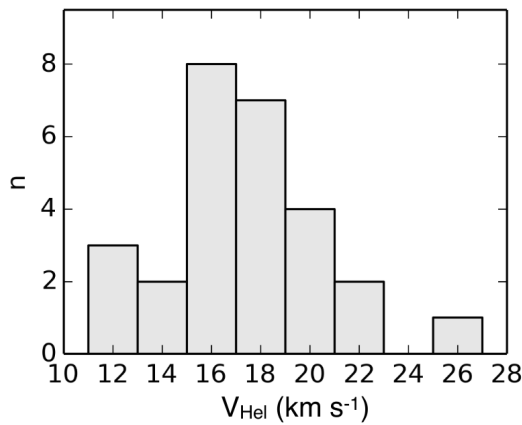}
\caption{ Histogram for the heliocentric velocities measured in the \ion{Li}{1} 6708 \AA~line observed from 2011B-2013B.
\label{fig:Li_velocities}}
\end{figure}

{
Of the absorption lines in Figure \ref{fig:CaI_LiI}, the \ion{Li}{1} line provides the best opportunity to investigate the line profiles as the other adjacent absorption lines are shallower. The absorption line profiles are symmetric with a FWHM of 40--60 km s$^{-1}$ for half of the spectra observed in 2011B--2013B. The remaining half show asymmetric line profiles (Figure \ref{fig:Li_asymmetries}). All the absorption line profiles in Figure \ref{fig:Li_asymmetries}, except those of 2012 September 26 and November 28, show an asymmetry near the bottom of the absorption line. This may be due to the non-uniform distribution of hot spots on the stellar surface. Several profiles in Figure \ref{fig:Li_asymmetries} show redshifted wing absorption due to an inflow, and that of 2014 Jan 17 shows blueshifted wing absorption due to a wind.

Figure \ref{fig:Li_averaged_profiles} shows the \ion{Li}{1} profiles averaged over 2011B--2013B and 2010B+2014B, respectively. The \ion{Li}{1} absorption in 2010B and 2014B is not clear in Figure \ref{fig:CaI_LiI}, but Figure \ref{fig:Li_averaged_profiles} clearly show the feature with an extra redshifted absorption at up to $\sim$100 km s$^{-1}$. Both profiles in Figure \ref{fig:Li_averaged_profiles} also show marginal blueshifted emission at $V_{\rm Hel} \sim -60$ km s$^{-1}$. In this context, the line profile averaged in 2010B+2014B may be an inverse P-Cygni profile.

We measure the radial velocity of the star in 2011B--2013B using \ion{Li}{1} absorption via Gaussian fitting, avoiding the blueshifted and redshifted wings (Figure \ref{fig:Li_velocities}). This yields a median heliocentric velocity of 17 km s$^{-1}$, agreeing with previous measurements of systemic velocities \citep[14--16 km s$^{-1}$, ][]{Petrov01a,Folha01}. The variation from the median heliocentric velocity is up to $\pm$5 km s$^{-1}$, which is similar to the periodic (2.8 days) variation of $\pm$6 km s$^{-1}$ observed by \citep{Petrov01a}.
}

 
\subsection{Emission Lines}

\begin{figure*}
\epsscale{2.1}
\plotone{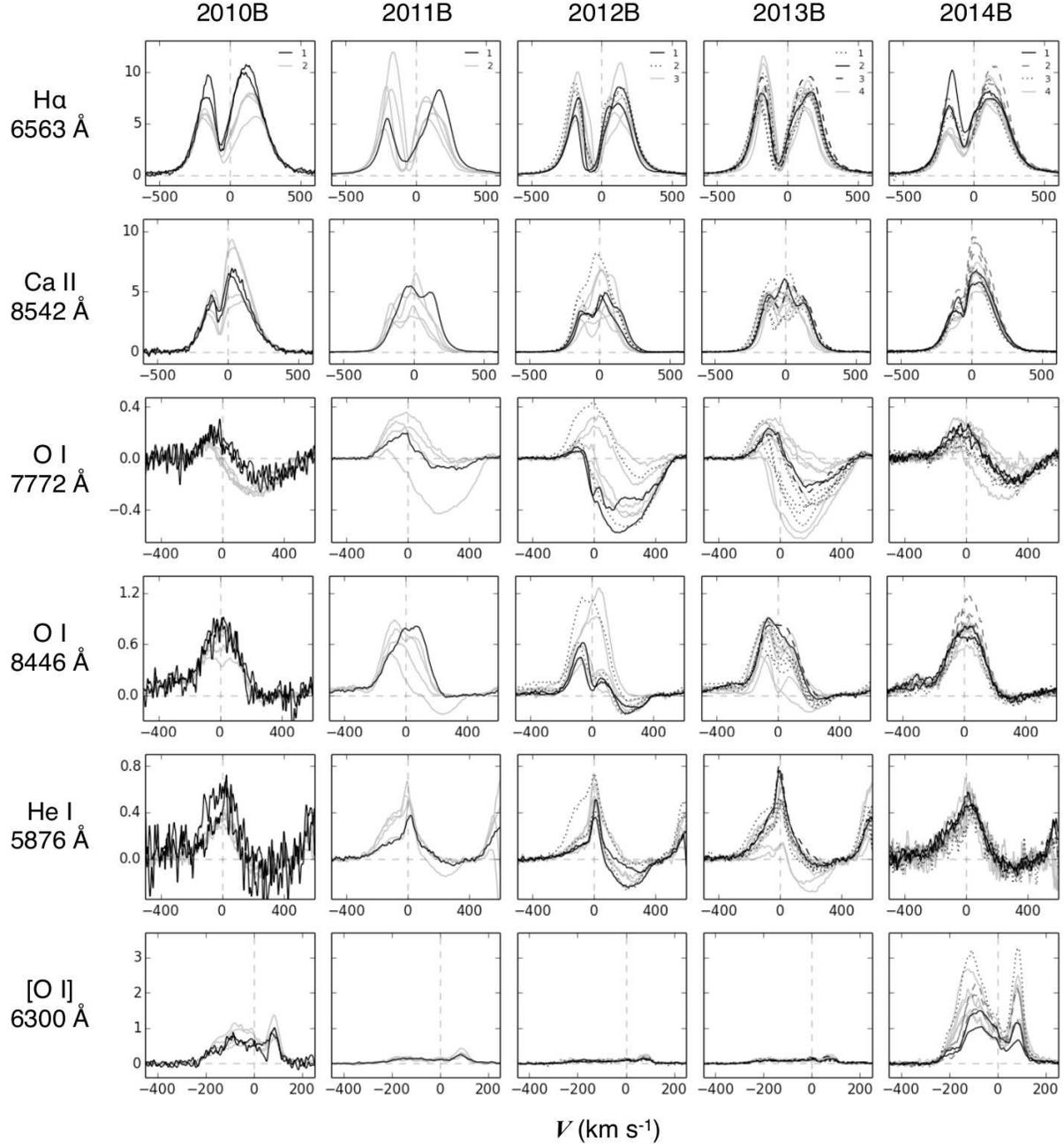}
\caption{The line profiles observed in the individual semesters. Each line profile is convolved with a Gaussian with a FWHM of 10 km s$^ {-1}$. The intensity is normalized to the continuum level.
Different styles (solid/dotted/dashed) and colors (black/gray) are used for different observing runs in each semester. See Table
\ref{table:log} for the details of the semesters, observing runs and dates.
\label{fig:profiles1}}
\end{figure*}

\begin{figure*}
\epsscale{2.1}
\plotone{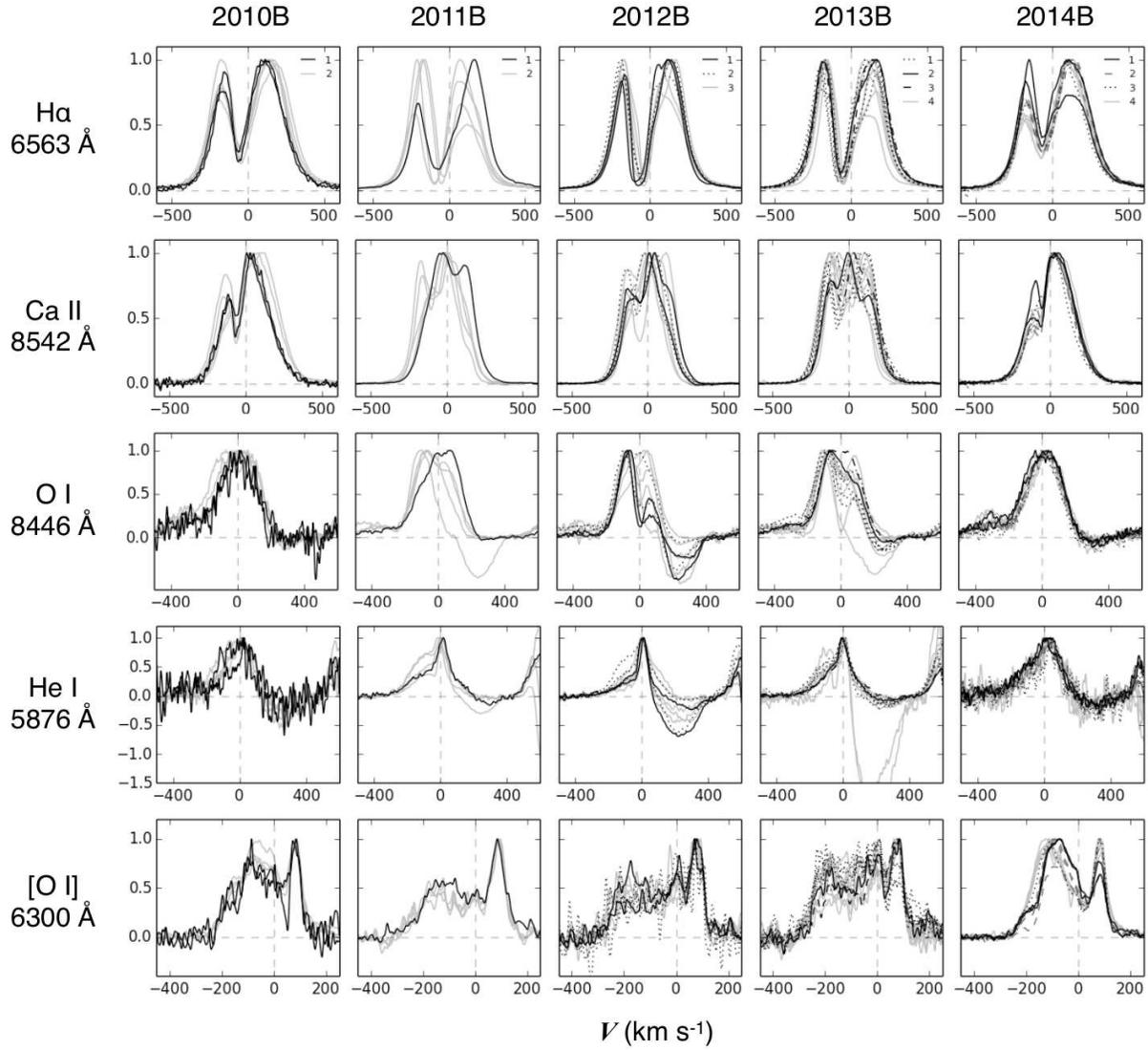}
\caption{Same as Figure \ref{fig:profiles1} but each line profile is normalized to the peak intensity.
Those of \ion{O}{+1} 7772 \AA~are not shown as the emission component is marginal, and the above
intensity scaling over-enhance the redshifted absorption, making comparisons difficult.
\label{fig:profiles2}}
\end{figure*}

Figures \ref{fig:profiles1} and \ref{fig:profiles2} show the line profiles for H$\alpha$ 6563 \AA, \ion{Ca}{+2} 8542 \AA, \ion{O}{+1} 7772 \AA~and 8446 \AA, \ion{He}{+1} 5876 \AA, and [\ion{O}{+1}] 6300 \AA. We adopt a stellar heliocentric velocity of { 17 km s$^{-1}$ based on the measurements in Section 3.1.}
In Figure \ref{fig:profiles1} the observed intensity is normalized to the continuum. In Figure \ref{fig:profiles2} we normalize the profiles shown in Figure \ref{fig:profiles1} by the peak intensity to better clarify changes in line profiles. 
Following \citet{Muzerolle98_opt} we have removed photospheric absorption in these profiles (except H$\alpha$ and \ion{Ca}{+2} lines) by scaling the spectrum of the WTTS Par 1379, which provided the best fit in Section 3.1, and subtracting it. The scaling factor was determined using a few of the deepest photospheric lines in the adjacent continuum. After this process we convolved each profile using a Gaussian with FWHM of 10 km s$^{-1}$ to improve the signal-to-noise. 

These profiles are similar to those previously observed \citep[e.g.,][]{Mundt84,Hamann92a,Fernandez95,Muzerolle98_opt,Petrov01a,Beristain01,Alencar05,Petrov15}. However, they show differences between photometric bright (2011B--2013B) and faint periods (2010B, 2014B), as described below in detail.

All of the H$\alpha$ { emission line} profiles are similar in the following characteristics: a FWHM and a full-width zero intensity (FWZI) of $\sim$500 and $\sim$1000 km s$^{-1}$, respectively; blue and redshifted peaks at $\sim -200$ to $-150$ and $\sim$50 to 150 km s$^{-1}$, respectively; and a minimum intensity at the middle of the line profile at $\sim - 50$ km s$^{-1}$. The intensity minimum at $\sim -50$ km s$^{-1}$ almost corresponds to the continuum level in 2010B--2013B, while it shows about 30--40 \% of the peak intensity in 2010B and 2014B.

\begin{figure}
\epsscale{0.9}
\plotone{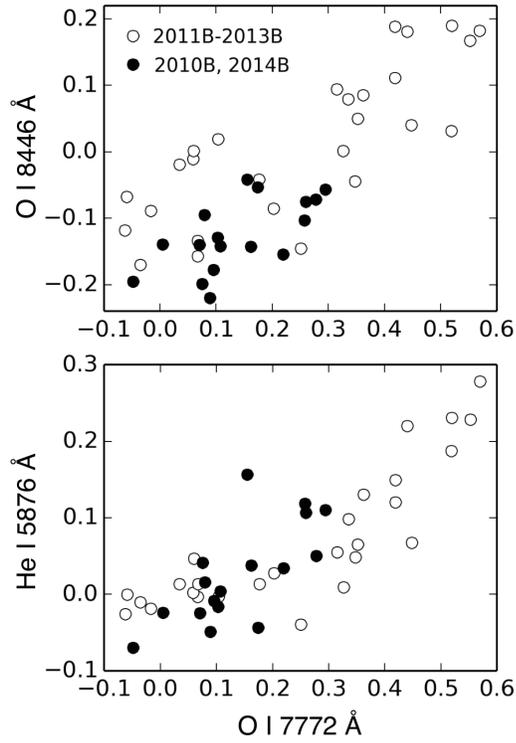}
\caption{($left$) Correlations of the redshifted absorption (averaged at 150--200 km s$^{-1}$, normalized to the continuum)
between the \ion{O}{+1} 7772 \AA, \ion{O}{+1} 8446 \AA, and \ion{He}{+1} 5876 \AA~ lines. The open circles are values for
2011B--2013B; the filled circles are for 2010B and 2014B.
\label{fig:correlations}}
\end{figure}

The \ion{Ca}{+2}, \ion{O}{+1}, and \ion{He}{+1} lines show more notable differences between the photometrically bright (2011B--2013B) and faint periods (2010B, 2014B). This difference is most remarkable in the \ion{Ca}{+2} 8542 \AA~{ emission} line. The line profiles in 2010B and 2014B exhibit blueshifted and redshifted peaks at $\sim$--100 and $\sim$50 km s$^{-1}$, respectively, with the redshifted peak brighter than the blueshited peak by typically $\sim$30 \%. In contrast, the line profiles in 2011B--2013B show complicated variations in each semester, exhibiting 1-3 peaks at --150 to 120 km s$^{-1}$.
The \ion{O}{+1} 7772 \AA~and 8446 \AA, and \ion{He}{+1} { emission is associated with} redshifted absorption at up to $\sim 500$ km s$^{-1}$, and the time variation of the absorption appears to be larger in 2011B--2013B than 2010B and 2014B. The absorption depths between these three lines are well correlated as shown in  Figure \ref{fig:correlations}.

\begin{figure*}
\epsscale{2.2}
\plotone{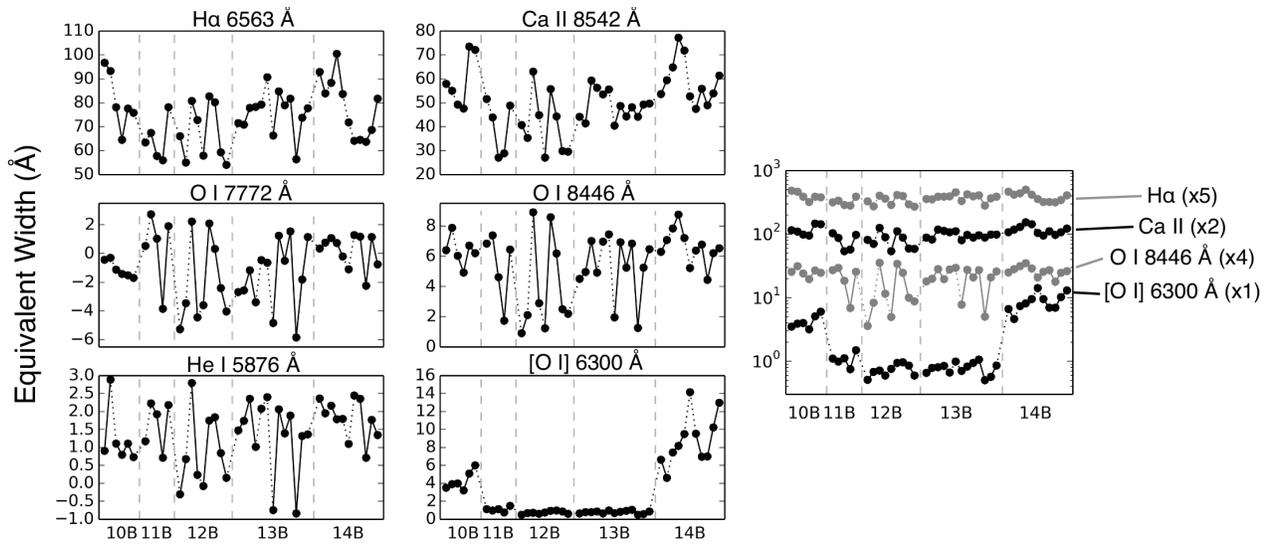}
\caption{($left$) Time variation of the equivalent widths of the emission lines. The values for different visits are plotted with a constant interval along the horizontal axis regardless of the actual time intervals. The vertical axes are shown
in linear scale. The values observed in the same observing run are connected using sold lines. The values
between different observing runs are connected using dotted lines.
($right$) Same but for H$\alpha$, \ion{Ca}{+2} 8542 \AA, \ion{O}{+1} 8446 \AA, and [\ion{O}{+1}] 6300 \AA~
(i.e. the lines with only positive equivalent widths during our observations) in a single plot.
The vertical axis are shown in log scale to compare relative variation between different nights. All but 
[\ion{O}{+1}] 6300 \AA~are scaled with scaling factors shown in the figure.
\label{fig:EWs}}
\end{figure*}

The line-to-continuum ratio of the [\ion{O}{+1}] 6300 \AA~{ emission} line is remarkably large in the photometrically faint periods (2010B, 2014B), but small in the bright period (2011B--2013B). The peak line-to-continuum ratio is 0.7--1.4, 0.2--0.4, 0.1--0.3, 0.1--0.3, and 1.0--3.3 in 2010B, 2011B, 2012B, 2013B, and 2014B, respectively. Figure \ref{fig:EWs} shows the time variation of the equivalent widths for the seven lines. The [\ion{O}{+1}] 6300 \AA~{ emission} shows a decrease in equivalent width during 2011B-2013B by a factor of $\sim$10. This decrease is primarily attributed to different continuum fluxes at photometrically bright ($V$ 10--11 mag) and faint ($V \sim 12.5$ mag. at the minima) periods \citep{Rodriguez13,Antipin15,Petrov15}. In contrast, the equivalent widths of the other lines were relatively similar for the whole period of observations.

In Figure \ref{fig:EWs} the \ion{O}{+1} 7772 and 8446 \AA, and \ion{He}{+1} 5876 \AA~ lines show a relatively large variation over 2011B--2013B compared to 2010B and 2014B. In Table \ref{table:EWs} we summarize the variation of the equivalent widths in the individual semesters.
The variation in the \ion{He}{+1} 5876 \AA~ line in 2011B is comparable to 2010B and 2014B but significantly larger in 2012B and 2013B. A large variation in equivalent widths of the \ion{O}{+1} and \ion{He}{+1} lines in 2011B--2013B is attributed to the large variation of redshifted absorption during this period (Figures \ref{fig:profiles1} and \ref{fig:profiles2}).

\begin{figure*}
\epsscale{2.2}
\plotone{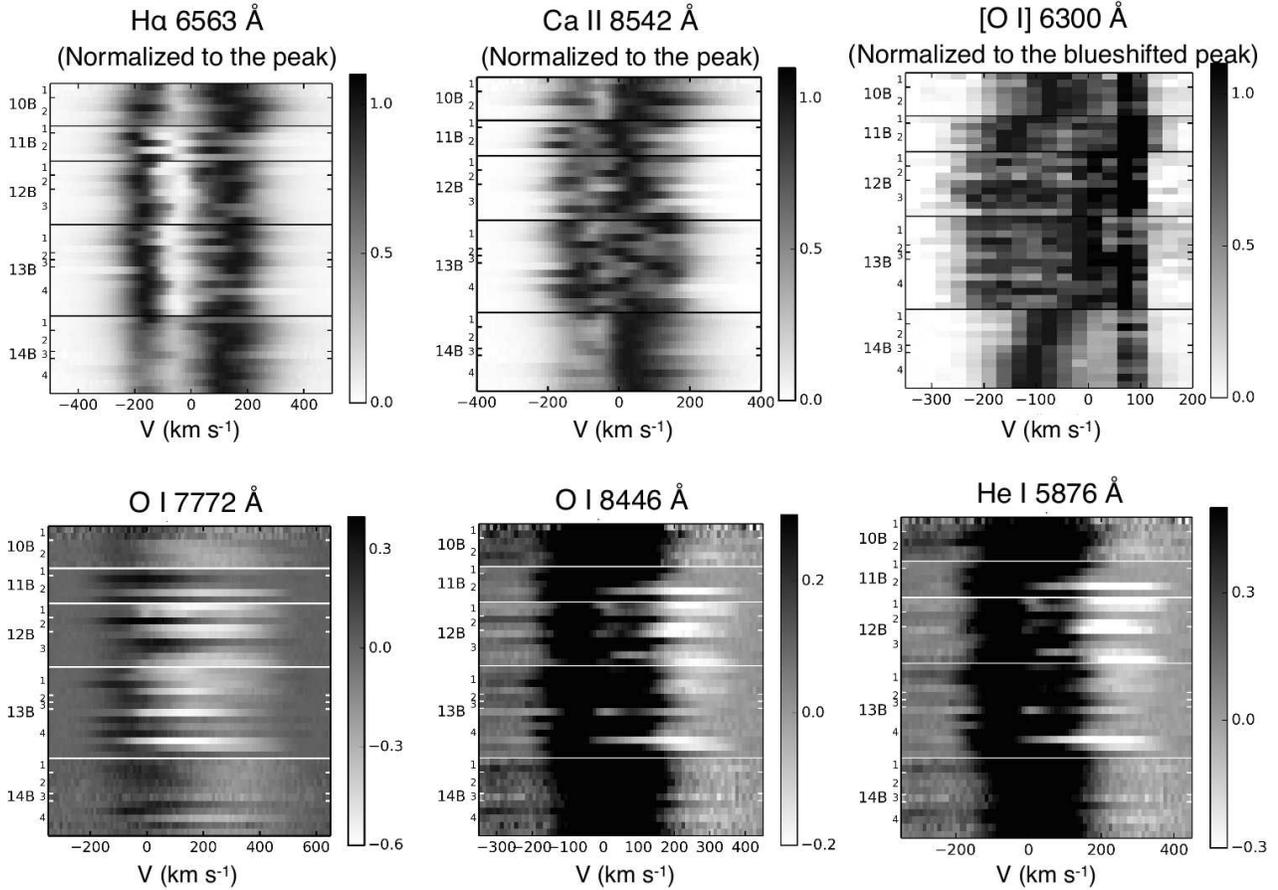}
\caption{
The line intensities as a function of velocity (horizontal axis) and time (vertical axis). The semesters and runs (see Table \ref{table:log} for details) are labeled at the left side of each figure. The line profiles normalized to the peak intensity (Figure \ref{fig:profiles2}) and the continuum (Figure \ref{fig:profiles1}) are used for the upper and lower figures, respectively. The contrasts for the upper figures are adjusted to clarify variation in the emission features, while those for the lower figures are adjusted to clarify variation in the redshifted absorption. Each line profile is binned for every 30 km s$^{-1}$ for the [\ion{O}{+1}] 6300 \AA~line, and every 10 km s$^{-1}$ for the others.
\label{fig:gray}}
\end{figure*}

Figure \ref{fig:gray} shows line intensities as a function of velocity (the horizontal axis) and time (the vertical axis) to better investigate the time variation of emission and absorption features. In the photometrically bright period (2011B--2013B), complicated variations in the \ion{Ca}{+2} 8542 \AA~ { emission} profiles and large variations in redshifted absorption in the \ion{O}{+1} and  \ion{He}{+1} lines occurred even between the nearest visits, within a few days. In contrast, the H$\alpha$ { emission} line profiles are significantly more stable, as also shown in Figure \ref{fig:profiles2}.

In the [\ion{O}{+1}] 6300 \AA~{ emission} the peak velocity of the redshifted emission appears to remain constant over the entire period of the observations, as shown in Figures \ref{fig:profiles2} and \ref{fig:gray}. In contrast, the blueshifted emission appears to have long-term ($>$1 yr) variation in kinematics. However, higher signal-to-noise would be required for the photometrically bright period (2011B--2013B) for confirmation. In 2014B the peak velocity of the blueshifted emission changed from --80 to --120 km s$^{-1}$, exhibiting a relatively large change between September and November 2014 (Semester 2 and 3 in Figure \ref{fig:gray} and Table \ref{table:log}).


\section{DISCUSSION}


\subsection{Implications for Variability in the Continuum and Emission Lines}

{ 
Optical and near-infrared photometry, and X-ray spectrophtometry showed remarkable flux decreases (by 2--3 mag.) in 2010 and 2014 for RW Aur A \citep{Antipin15,Schneider15} and the A+B binary system \citep{Rodriguez13,Petrov15,Shenavrin15}. \citet{Shenavrin15} also show that the fluxes at $J$-, $H$-, and $K$-bands (1.25, 1.65 and 2.2 \micron, respectively) decreased but those at $L$- and $M$-bands (3.5 and 4.5 \micron, respectively) increased in 2014--2015.

Our spectra show time variations related to the above flux changes. First, photospheric absorption is clearly observed in the bright period (2011B--2013B), but is absent (or marginal) in the faint periods (2010B and 2014B). Secondly, emission and absorption line profiles show different variabilities between the bright and faint periods. In the bright period the \ion{Ca}{+2} 8542 \AA~emission profiles changed dramatically, and the redshifted absorption in the \ion{O}{+1} and \ion{He}{+1} lines showed a large variation. These contrast to relatively stable line profiles in the faint periods.

In this subsection we discuss two scenarios for the observed variabilities: (1) occultations of the star without a binary companion by dusty disk or fragments; (2) occultations in an unresolved binary system; and (3) changes in the activity of mass accretion. None of the scenarios can simply explain all the observed trends, and more theoretical work is needed to further investigate the feasibilities of scenarios (1) and (3).

\subsubsection{Occultations without a binary companion}


\cite{Rodriguez13,Antipin15,Petrov15,Schneider15,Shenavrin15} attribute the decrease of fluxes to occultation by a dusty circumstellar material, either fragments due to tidal interaction of the RW Aur A disk with RW Aur B \citep{Rodriguez13,Schneider15} or a dusty wind \citep{Shenavrin15,Petrov15}. The presence of tidal interaction between the RW Aur A disk and RW Aur B was first discussed by \citet{Cabrit06} to explain a tidal tail observed in millimeter CO emission. \citet{Dai15} conducted numerical simulations and succeeded in reproducing its major morphological and kinematic features. These authors show that this tidal interaction may create fragments of clouds which pass in front of RW Aur A. 

No substantial reddening was observed during the faint periods \citep[][; Section 3.1]{Petrov15,Antipin15,Schneider15}, and \citet{Petrov15,Schneider15} attributed this to large grains ($r\gg1$ \micron) in the occulter. \citet{Shenavrin15} reported an increase of the 3-5 \micron~fluxes in 2014--2015. These authors attributed this increase to hot ($\sim 1000$ K) dust thermal continuum associated with the occulter. If this is the case, the occulter must be located close to the star ($\sim$0.1 AU).

However, the occultation scenario does not simply explain the spectral variations shown in Section 3. The spectra should not change if the star, the accretion flows and a wind are uniformly occulted. The hot spots on the stellar surface, the emission accretion flows and a wind are not uniform, therefore the spectral variations would occur if (1) the occulter allows only a part of the stellar surface or the inflow/outflow to be observed; or (2) while the direct fluxes from the star and the inflow/outflow are fully occulted, some emission is still observed via scattering from circumstellar dust. However, it is not clear if these can explain the similar H$\alpha$ profiles and the H$\alpha$, \ion{Ca}{2}, \ion{O}{1} and \ion{He}{1} equivalent widths through the entire periods of observations.

Furthermore, the occultation scenarios have the following problems. The occultation scenario without scattering requires a sharp edge of the occulter, comparable to the stellar radii. The presence of the hot dust continuum suggests a temperature of $\sim 1000$ K. Gas with such a temperature would immediately expand well beyond the scale of a few stellar radii unless the structure is sustained by strong gravity or magnetic fields. The occultation scenarios proposed so far is not sufficiently detailed to discuss this issue. The occultation scenario with scattering would cause changes in the color of the optical continuum \citep{Clampin03,Ardila07,Wisniewski08}, however, such a color change was not observed for the faint periods \citep[Section 3.1]{Petrov15,Antipin15}.
}

\subsubsection{Occultation in an unresolved binary system}

{ 
One might think that the observed trends could be explained if RW Aur A is an unresolved binary. The observed fluxes toward RW Aur A flux would be dominated by the bright component in the bright period, but would be obscured during the faint periods, and as a result, the observed fluxes during these period would be dominated by the companion.

However, the facts below rather favour the scenario where the observed spectra are associated with the same star during the entire period. This scenario requires a significantly different brightness for the two binary components, and therefore the masses, spectral types and colors of the two stars would also be significantly different. However, a significant color change was not observed in the optical continuum \citep[Section 3.1]{Petrov15,Antipin15}. Furthermore, the shape of the line profiles and equivalent widths for H$\alpha$, \ion{O}{+1}, and \ion{He}{+1} lines are also very similar during the entire period of the observations.
}

\subsubsection{ Time-variable mass accretion?}

The observed variations in line profiles may be alternatively explained by unstable and stable regimes of magnetospheric mass accretion. Possible mechanisms for the variation of kinematics, and thereby line profiles, include a magnetic Rayleigh-Taylor (RT) instability \citep[e.g.,][]{Romanova08,Kurosawa13} or a magneto-rotational instability in the accretion disk \citep[e.g.,][]{Hawley00,Stone00_PPIV,Romanova12}. Of these mechanisms, the observed trends may particularly favor the RT scenario, as discussed below.

\citet{Romanova08,Kulkarni08} present 3-D magnetohydrodynamical simulations of magnetospheric mass accretion through the RT instability. These simulations show that unstable and stable mass accretion occur at high and low mass accretion rates, if the star has a dipole field misaligned relative to the rotation axis by a small angle ($< 30^\circ$). In the unstable regime, the star accretes through several transient elongated inflows developed by the RT instability. These inflows (``tongues'') deposit gas at random places on the stellar surface. Their gas kinematics changes even within a few rotational periods of the star. In contrast, the star accretes in ordered funnel streams in the stable regime.

\citet{Kurosawa13} carried out radiative transfer simulations for the unstable and stable regimes of magnetospheric accretion, and found complicated time variation of the Balmer lines profiles similar to those seen in the \ion{Ca}{+2} lines in Figures \ref{fig:profiles1} and \ref{fig:profiles2}. In contrast, the H$\alpha$ profiles observed for RW Aur A do not show such a complicated variation. This may be due to a different origin for the emission line, e.g., from a wind \citep[e.g.,][]{Najita00,Takami03,Alencar05,Kurosawa12}.

In summary, the RT models by \citet{Romanova08,Kulkarni08,Kurosawa13} explain the bright and faint periods of RW Aur as follows. The bright periods would be due to high mass accretion rates, which cause a relatively bright blue excess continuum \citep[e.g.][for a review]{Calvet00}. The high accretion rates would also cause unstable magnetospheric accretion flows, which cause a complicated time variation in the \ion{Ca}{+2} line profiles and a large variation of redshifted absorption in \ion{O}{+1} and \ion{He}{+1} lines. On the other hand, the faint periods are due to low mass accretion rates, for which we expect lower optical continuum flux and more stable accretion flow.

\citet{Kurosawa13} also showed the presence of a relatively large and periodic redshifted absorption in Balmer line profiles in the stable periods. In contrast, the observed variation of redshifted absorption in \ion{O}{+1} and \ion{He}{+1} lines are significantly smaller in the faint stable periods than the bright unstable period. Such a discrepancy may be attributed to either (1) different mass accretion rates; (2) different excitation conditions; (3) a combination of the inflow geometry and a viewing angle; or (4) a more azimuthally uniform inflow. Observations of the spectra with full stellar rotational periods will allow us to further to test the validity of the RT models.

{ However, it is not clear if the time variable mass accretion can explain the following trends and features in the faint periods: (1) the absence of photospheric absorption; (2)  redshifted absorption in the \ion{Li}{1} line; and (3) hot thermal dust emission.}

\subsection{A Physical Link Between Mass Accretion and Ejection}
Many young stellar objects including CTTSs are known to host a collimated jet \citep[e.g.,][for reviews]{Eisloffel00,Ray07,Frank14}. Understanding their driving mechanism and their physical link with protostellar evolution are also important issues in star formation theories. CTTSs have offered the best opportunity to observationally investigate this link because of their optically visible nature. Their line luminosities and the veiling toward CTTSs strongly support the theories which predict that the jets are powered by disk accretion \citep[e.g.,][]{Cabrit90,Hartigan95,Calvet97}.

The proposed theories of jet driving fall into the following two categories. The X-wind \citep[][]{Shu00} and magnetospheric ejection theories \citep[e.g.,][]{Hayashi96,Goodson99,Zanni13} predict that the jet launches from the stellar magnetosphere, or
the inner edge of the disk associated with the stellar magnetosphere. In contrast, the disk wind theories predict that the jet launching region covers a wider disk surface \citep[e.g.,][for a review]{Pudritz07}. However, observational studies of the jet launching have been hampered by the following issues. The angular resolutions of most present telescopes are far from sufficient for resolving the jet launching or flow accelerating regions, which are predicted to be within a few AU from the star. Such observations require an angular resolution of several milliarcseonds for a typical distance to nearby star forming regions ($\sim$140 pc). Such resolutions could be achieved by some optical or near-infrared interferometers, but their spatial information is still limited.

Alternatively, simultaneous monitoring of jet structures and signatures of magnetospheric accretion would allow us to test theories of jet driving and mass accretion \citep[][]{Chou13}. \citet{Lopez03} measured the proper motions of the RW Aur jets for which the ejection has occurred during the past $\sim$70 years. These suggest time variable mass ejection with at least two modes: one irregular and asymmetric (possibly random) on timescales of $\le$3--10 yr, and another more regular with a $\sim$20 year period. Variabilities shown in recent photometry \citep{Petrov15} and our high resolution spectroscopy show a similar time scale ($\sim$4 yr between the faint periods in 2010 and 2014).

Monitoring observations of the jet structures together with { optical spectra} in our study may be useful for understanding whether the jets are launched near the stellar magnetosphere, and how the time variation of mass accretion affects the mass ejection seen in the jets. 
{ If we find a clear correlation between these, it would also imply that the time variations of the optical spectra discussed in this work are due to changes in mass accretion activity.}

\section{SUMMARY}

We present monitoring observations of the active T Tauri star RW Aur using optical high-resolution spectroscopy with CFHT-ESPaDOnS. The observations were made from 2010 October to 2015 January in the autumn-winter semesters (August--January),  with 2--4 observing runs for each semester, and 1--7 visits during each run, with intervals of 1--10 nights, with a spectral resolution of $R$=68,000. Optical photometry in the literature shows bright, stable fluxes over most of this period, with lower fluxes (by 2-3 mag.) in 2010 and 2014.

In the bright period (semesters 2011B--2013B, August 2011 to January 2014) our spectra show clear photospheric absorption, whose degree of variation ($r$=0.7--5 at 6000 \AA) and color ($F_{\lambda 5800} / F_{\lambda 8635}$) agrees with standard veiling theory. In contrast, photospheric absorption is absent or marginal in 2010B (October--November 2010) and 2014B (August 2014 to January 2015), which correspond to the faint periods.

Different trends of time variation are also observed in  \ion{Ca}{+2} 8542 \AA, \ion{O}{+1} 7772 and 8446 \AA, and \ion{He}{+1} 5876 \AA~profiles between the bright and faint periods. In the bright period the \ion{Ca}{+2} { emission} profiles show complicated variation, and redshifted absorption in the \ion{O}{+1} and \ion{He}{+1} lines show a large variation. In contrast, the line profiles are similar to each other during the faint periods. 

{ 
We have discussed the following possibilities to attempt to explain the variations in the optical spectra and optical, near-IR and X-ray fluxes: (1) occultation without a binary companion; (2) occultation in an unresolved binary; and (3) changes in mass accretion. None of the scenarios can simply explain all the observed trends. More theoretical work would be able to justify either of scenario (1) or (3). Alternatively, simultaneous observations of the jet structures would allow us to investigate if the observed time variation is due to mass accretion.
}

\acknowledgments
{ We thank the anonymous referee for useful comments, and Drs. Silvia Alencar, Sylvie Cabrit, Catherine Dougados, Jerome Bouvier and Ronald Taam for useful discussion.}
M.T. and Y.J.W are supported from Ministry of Science and Technology (MoST) of Taiwan (Grant No. 103-2112-M-001-029, 104-2119-M-001-018).
{ R.G.M. acknowledges support from program UNAM-DGAPA-PAPIIT IA101715.}
This research made use of the Simbad database operated at CDS, Strasbourg, France, and the NASA's Astrophysics Data System Abstract Service.

{\it Facilities:} \facility{Canada-France-Hawaii Telescope (ESPaDOnS)}.



\clearpage




\clearpage

\clearpage

\clearpage

\clearpage

\clearpage

\clearpage

\clearpage

\clearpage

\clearpage

\begin{table*}
\caption{Stellar Properties \label{table:RWAur}}
\begin{tabular}{lcc}
\tableline\tableline
Parameter								& Value 		& References\tablenotemark{a}	\\ \tableline
Distance (pc)							& $140\pm 20$		& 1	\\
Spectral Type							& K1--K4			& 2, 3, 4, 5	\\
Stellar Mass ($M_\odot$) 					& $1.3 \pm 0.2$ 	& 4, 6\\
log (Age/yr)							& $6.9 \pm 0.3$	& 4\\
log ({\it \.{M}}$/M_\odot$ yr$^{-1}$)				& --7.5			& 4\\
\tableline
\end{tabular} \\
\tablenotetext{a}{
(1) \citet{Bertout06} ;
(2) \citet{Hartigan95} ;
(3) \citet{Muzerolle98_opt} ;
(4) \citet{White01} ;
(5) \citet{Petrov01a} ;
(6) \citet{Woitas01}}
\end{table*}

\begin{table}
\caption{Log of the observations \label{table:log}}
\begin{tabular}{lcl}
\tableline\tableline
Semester & Run & Dates (yyyy-mm-dd)\\ \tableline
2010B	& 1	&	2010-10-(16, 21)	\\
		& 2	&	2010-11-(17, 21, 25, 27)	\\
2011B	& 1	&	2011-08-20		\\
		& 2	&	2012-01-(05, 09, 11\tablenotemark{a}, 15\tablenotemark{a})	\\
2012B	& 1	&	2012-09-(26\tablenotemark{a}, 29)	\\
		& 2	& 	2012-11-(25, 28) ; 2012-12-(01, 08)	\\
		& 3	&	2012-12-(22, 25, 28) \\
2013B	& 1	&	2013-08-(15, 17, 21, 28) \\
		& 2 	&	2013-09-26 \\
		& 3	&	2013-11-23 \\
		& 4	&	2014-01-(10\tablenotemark{a}, 11, 15, 16, 17, 19, 20) \\
2014B	& 1	&	2014-08-(15, 19) \\
		& 2	&	2014-09-(04, 10, 15) \\
		& 3	&	2014-11-(05) \\
		& 4	&	2014-12-(20, 22, 29) ; 2015-01-(07, 11) \\		
\tableline
\end{tabular} \\
\tablenotetext{a}{Two exposures were obtained. See text for details.}
\end{table}

\begin{table}
\caption{Measured Veiling $r$ \label{table:veiling}}
\begin{tabular}{ccc}
\tableline\tableline
Date (yyyy-mm-dd)	& 5720--5860 \AA	& 6000--6130 \AA  \\ \tableline
2011-08-20		& 0.9				& 0.9 	\\
2012-01-05		& 0.7				& 0.7 	\\
2012-01-09		& 0.6				& 0.7		\\
2012-01-11		& 4.8				& 4.2		\\
2012-01-15		& 0.6				& 0.7		\\
2012-09-26		& 2.3				& 1.8		\\
2012-09-29		& 1.7				& 1.5		\\
2012-11-25		& 1.2				& 1.3		\\
2012-11-28		& 3.6				& 2.9		\\
2012-12-01		& 1.4				& 1.4		\\
2012-12-08		& 0.7				& 0.7		\\
2012-12-22		& 0.9				& 1.2		\\
2012-12-25		& 1.3				& 1.2		\\
2012-12-28		& 3.0				& 3.3		\\
2013-08-15		& 2.8				& 1.5		\\
2013-08-17		& 3.6				& 3.2		\\
2013-08-21		& 3.1				& 3.3		\\
2013-08-28		& 3.8				& 1.7		\\
2013-09-26		& 2.0				& 1.8		\\
2013-11-23		& 1.3				& 1.1		\\
2014-01-10		& 2.5				& 1.1		\\
2014-01-11		& 2.1				& 1.5		\\
2014-01-15		& 1.3				& 0.9		\\
2014-01-16		& 1.5				& 1.0		\\
2014-01-17		& 4.3				& 2.9		\\
2014-01-19		& 3.1				& 2.7		\\
2014-01-20		& 1.0				& 0.8	\\
\tableline
\end{tabular} \\
\end{table}

\clearpage

\begin{table*}
\caption{Ranges of Equivalent Widths\tablenotemark{a} (\AA) \label{table:EWs}}
{\scriptsize
\begin{tabular}{lccccc}
\tableline\tableline
Line 	& 2010B	& 2011B	& 2012B	& 2013B	& 2014B \\ \tableline
H$\alpha$ 6563 \AA		& 64.5 $\sim$ 96.8 (32.3)	& 56.0 $\sim$ 78.2 (22.2)	& 54.1 $\sim$ 82.7 (28.6)	& 56.4 $\sim$ 90.9 (34.3)	& 63.8 $\sim$ 100.5 (36.8)\\
\ion{Ca}{+2} 8542 \AA	& 47.6 $\sim$ 73.4 (25.9)	& 27.1 $\sim$ 51.7 (24.6)	& 27.1 $\sim$ 63.0 (35.9)	& 40.4 $\sim$ 59.3 (18.9)	& 47.5 $\sim$ 77.0 (29.7)	\\
\ion{O}{+1} 7772 \AA		& --1.7 $\sim$ --0.3 (1.4)	& --3.9 $\sim$ 2.7 (6.6)	& --5.3 $\sim$ 2.2 (7.5)	& --5.8 $\sim$ 1.5 (7.4)	& --2.2 $\sim$ 1.3 (3.5)	\\
\ion{O}{+1} 8446 \AA		& 4.9 $\sim$ 7.9 (3.0)	& 1.7 $\sim$ 7.4 (5.6)	& 0.9 $\sim$ 8.9 (8.0)	& 1.3 $\sim$ 7.4 (6.2)	& 4.4 $\sim$ 8.8 (4.3)	\\
\ion{He}{+1} 5876 \AA	& 0.7 $\sim$ 2.9 (2.2)	& 0.7 $\sim$ 2.2 (1.5)	& --0.3 $\sim$ 2.8 (3.1)	& --0.8 $\sim$ 2.4 (3.2)	& 0.7 $\sim$ 2.4 (1.7)	\\
$[$\ion{O}{+1}$]$ 6300 \AA 
& 3.2 $\sim$ 6.0 (2.8)	& 0.8 $\sim$ 1.5 (0.7)	& 0.5 $\sim$ 1.0 (0.5)	& 0.5 $\sim$ 1.1 (0.6)	& 4.6 $\sim$ 14.1 (9.5)	\\
\tableline
\end{tabular} \\
}
\tablenotetext{a}{The values in the brackets are differences between the minimum and maximum values.}
\end{table*}


\end{document}